\documentstyle[wstwocl]{article}
\begin{document}

\bibliographystyle{unsrt}    % for BibTeX - sorted numerical labels

\newcommand{\st}{\scriptstyle}
\newcommand{\sst}{\scriptscriptstyle}
\newcommand{\mco}{\multicolumn}
\newcommand{\epp}{\epsilon^{\prime}}
\newcommand{\vep}{\varepsilon}
\newcommand{\ra}{\rightarrow}
\newcommand{\ppg}{\pi^+\pi^-\gamma}
\newcommand{\vp}{{\bf p}}
\newcommand{\ko}{K^0}
\newcommand{\kb}{\bar{K^0}}
\newcommand{\al}{\alpha}
\newcommand{\ab}{\bar{\alpha}}
\def\be{\begin{equation}}
\def\ee{\end{equation}}
\def\bea{\begin{eqnarray}}
\def\eea{\end{eqnarray}}
\def\CPbar{\hbox{{\rm CP}\hskip-1.80em{/}}}%temp replacement due to no font

\def\ap#1#2#3   {{\em Ann. Phys. (NY)} {\bf#1} (#2) #3.}
\def\apj#1#2#3  {{\em Astrophys. J.} {\bf#1} (#2) #3.}
\def\apjl#1#2#3 {{\em Astrophys. J. Lett.} {\bf#1} (#2) #3.}
\def\app#1#2#3  {{\em Acta. Phys. Pol.} {\bf#1} (#2) #3.}
\def\ar#1#2#3   {{\em Ann. Rev. Nucl. Part. Sci.} {\bf#1} (#2) #3.}
\def\cpc#1#2#3  {{\em Computer Phys. Comm.} {\bf#1} (#2) #3.}
\def\err#1#2#3  {{\it Erratum} {\bf#1} (#2) #3.}
\def\ib#1#2#3   {{\it ibid.} {\bf#1} (#2) #3.}
\def\jmp#1#2#3  {{\em J. Math. Phys.} {\bf#1} (#2) #3.}
\def\ijmp#1#2#3 {{\em Int. J. Mod. Phys.} {\bf#1} (#2) #3.}
\def\jetp#1#2#3 {{\em JETP Lett.} {\bf#1} (#2) #3.}
\def\jpg#1#2#3  {{\em J. Phys. G.} {\bf#1} (#2) #3.}
\def\mpl#1#2#3  {{\em Mod. Phys. Lett.} {\bf#1} (#2) #3.}
\def\nat#1#2#3  {{\em Nature (London)} {\bf#1} (#2) #3.}
\def\nc#1#2#3   {{\em Nuovo Cim.} {\bf#1} (#2) #3.}
\def\nim#1#2#3  {{\em Nucl. Instr. Meth.} {\bf#1} (#2) #3.}
\def\np#1#2#3   {{\em Nucl. Phys.} {\bf#1} (#2) #3.}
\def\pcps#1#2#3 {{\em Proc. Cam. Phil. Soc.} {\bf#1} (#2) #3.}
\def\pl#1#2#3   {{\em Phys. Lett.} {\bf#1} (#2) #3.}
\def\prep#1#2#3 {{\em Phys. Rep.} {\bf#1} (#2) #3.}
\def\prev#1#2#3 {{\em Phys. Rev.} {\bf#1} (#2) #3.}
\def\prl#1#2#3  {{\em Phys. Rev. Lett.} {\bf#1} (#2) #3.}
\def\prs#1#2#3  {{\em Proc. Roy. Soc.} {\bf#1} (#2) #3.}
\def\ptp#1#2#3  {{\em Prog. Th. Phys.} {\bf#1} (#2) #3.}
\def\ps#1#2#3   {{\em Physica Scripta} {\bf#1} (#2) #3.}
\def\rmp#1#2#3  {{\em Rev. Mod. Phys.} {\bf#1} (#2) #3.}
\def\rpp#1#2#3  {{\em Rep. Prog. Phys.} {\bf#1} (#2) #3.}
\def\sjnp#1#2#3 {{\em Sov. J. Nucl. Phys.} {\bf#1} (#2) #3.}
\def\spj#1#2#3  {{\em Sov. Phys. JEPT} {\bf#1} (#2) #3.}
\def\spu#1#2#3  {{\em Sov. Phys.-Usp.} {\bf#1} (#2) #3.}
\def\zp#1#2#3   {{\em Zeit. Phys.} {\bf#1} (#2) #3.}

\setcounter{secnumdepth}{2} % Number sections and subsections

%%%%%%%%%%%%%%%%%%%%%%%%%%%%%%%%%%%%%%%%%%%%%%%%%%
%                                                %
%    BEGINNING OF TEXT                           %
%                                                %
%%%%%%%%%%%%%%%%%%%%%%%%%%%%%%%%%%%%%%%%%%%%%%%%%%

\title{NEW BOUNDS ON R-PARITY VIOLATING COUPLINGS}

\firstauthors{Marc Sher }

\firstaddress{Physics Department, College of William and Mary, \\
Wiliamsburg, VA  23187 USA}

%\secondauthors{ }

%if there are no second authors then comment out the line and adjust the
%maketitle command and \def\secondauthor... command in snow.sty

%\secondaddress{}

%if there are no second authors then comment out the line and adjust the
%maketitle command and \def\secondaddress... command in snow.sty

\twocolumn[\maketitle\abstracts{}]

\def\beq{beginequation}
\def\eeq{endequation}
\def\lm{\lambda}
\def\lusd{\lm_{usd}''}
\def\lubd{\lm_{ubd}''}
\def\lubs{\lm_{ubs}''}
\def\lcsd{\lm_{csd}''}
\def\lcbd{\lm_{cbd}''}
\def\lcbs{\lm_{cbs}''}
\def\ltsd{\lm_{tsd}''}
\def\ltbd{\lm_{tbd}''}
\def\ltbs{\lm_{tbs}''}

Though the minimal supersymmetric standard model (MSSM)~\cite{one}
is a leading candidate for new physics beyond the standard model,
the conservation of R-parity, $R_p$, which is assumed in the model
has no real theoretical justification.  This has motivated many
authors~\cite{two} to consider alternatives in which $R_p$ is
explicitly broken.  In such models, sparticles can decay into
non-supersymmetric particles alone, leading to novel signatures in
search experiments and unusual decay processes.

The most general $R_p$-violating superpotential that one can write
with the MSSM superfields, in the usual notation, is
\be
W=\lm_{ijk}L_iL_j\bar{E}_k+\lm_{ijk}'L_iQ_j\bar{D}_k
+\lm_{ijk}''\bar{U}_i\bar{D}_j\bar{D}_k
\ee
Here, $i,j,k$ are generation indices and we have rotated away a
term of the form $\mu_{ij}L_iH_j$.  Since the $\lm_{ijk}$ term
is symmetric under exchange of $i$ and $j$, and antisymmetric in
color, it must be antisymmetric in flavor, thus we have
$\lm_{ijk}=-\lm_{jik}$.  Similarly, $\lm_{ijk}''=-\lm_{ikj}$.  The
number of couplings is then 36 lepton nonconserving couplings (9
of the $\lm$ type and 27 of the $\lm'$ type) and 9 baryon
nonconserving couplings (all of the $\lm''$ type) in total.

It is generally thought that $\lm$, $\lm'$ type couplings cannot
coexist with $\lm''$ type couplings since both baryon and lepton
number violations would lead to too rapid a proton decay.  For this
reason, previous authors have considered either lepton
nonconserving {\it or} baryon nonconserving couplings, but not
both.  We will first make this assumption and consider the baryon
number violating $\lm''$ couplings alone, since most of the
earlier effort has been focused on $\lm$ and $\lm'$
couplings~\cite{three}.  Later, we will examine proton decay in the
presence of all three types of couplings.

We can write the 9 different $\lm''$ couplings as $\ltbs, \ltbd,
\ltsd, \lcbs, \lcbd, \lcsd, \lubs, \lubd$ and $\lusd$.  Let us
first recount the existing constraints on these couplings.
 Brahmachari and Roy~\cite{br} showed that the requirement of
perturbative unification typically places a bound of between
$1.10$ and
$1.25$ on many of the couplings.  This was generalized to all the
couplings by Goity and Sher~\cite{goity}.  The latter also showed,
following earlier work~\cite{zw,barbieri},   that
$|\lusd|$  can be strongly bounded by the nonobservation of
double nucleon decay into two kaons (such as
${}^{16}O\rightarrow {}^{14}C\ K^+ K^+$, which would have been seen
in water Cerenkov detectors), and  $|\lubd|$  can be
strongly bounded by the nonobservation of neutron-antineutron
oscillations. Their bounds, for squark masses of 300 GeV, were
$|\lubd| < 5\times 10^{-3}$ and $|\lusd| < 10^{-6}$.  In the
work of Barbieri and Masiero~\cite{barbieri}, some bounds on
products of couplings were obtained by considering $K$-$\bar{K}$
mixing; these bounds will be discussed shortly.  Finally, bounds
from vertex corrections to the decay of the $Z$ into two
charged leptons and into $\bar{b}b$
 have recently been obtained~\cite{cern}; though
potentially interesting, with present data they are not
significantly better than the bound from perturbative unification.

In this Talk, I  note that many additional and interesting
bounds on the $\lm''$ couplings can be obtained by considering rare
two-body nonleptonic decays of B mesons.  Let us begin by considering the
implications for
$\lambda''$ couplings from such processes..  Since
lepton number is assumed to be conserved, only $\Delta B=0$ and
$\Delta B=2$ decays can occur. Thus any bounds will be on the
 products of two $\lambda''$ couplings.  Furthermore,
since any B-decay will change the number of ``b-flavors'' by one
unit, bounds from there  will apply to products of the form
$\lambda_{{u_i}bs}\lambda_{{u_j}sd}$ or
$\lambda_{{u_i}bd}\lambda_{{u_j}sd}$. We first consider
$\Delta B=0$ (baryon number conserving) decays which actually give
the best bounds and later comment on the $\Delta B=2$ processes.

The details of the calculation can be found in the paper of Carlson,
Roy and Sher~\cite{crs}.  There, it is shown that the best bounds
come from the decay $B^+\rightarrow \bar{K}^oK^+$.  This has an extremely
small rate in the Standard Model, being penguin-suppressed and
also reduced by the small CKM element $V_{ub}$ in the amplitude.
The dominant diagrams contributing to this process for nonzero
$\lm''$ couplings come from squark exchange. We find that
\be
{\rm B.R.}(B^+\rightarrow \bar{K}^oK^+)\simeq
1.97|\ltbs\ltsd|^2(m_W/m_{\tilde{t}})^4.\ee

On using the recent experimental upper bound~\cite{cleo} of $5\times
10^{-5}$ on the branching ratio, and noting that one can replace
the scalar top with a scalar charm or scalar up, we then have
\be
{|\lm_{qbs}''\lm_{qsd}''|m^2_W\over m^2_{\tilde{q}}}< 5
\times 10^{-3},\ee
for $q=t,c,u$.

One can repeat the calculation for the decay $B^+\rightarrow
\bar{K}^o\pi^+$ (or equivalently $B^-\rightarrow
{K}^o\pi^-$) in much the same way.  The result (with only the
t-squark contribution being considered) is
\be
{|\lm_{qbd}''\lm_{qsd}''|m^2_W\over m^2_{\tilde{q}}}< 4.1
\times 10^{-3},\ee
for $q=t,c,u$.

Additional bounds were obtained by Barbieri and
Masiero~\cite{barbieri} from the contribution of $K$-$\bar{K}$
mixing to the $K_L$-$K_S$ mass difference.  We have generalized
their results by including effects from all of the squarks, updated
top mass and CKM angles, etc., and find bounds on several combinations
of couplings (including the product $|\lm_{tbs}''\lm_{tbd}''|$), the
plot can be found in Ref. [9].

What about $\Delta B = 2$ decays?   One can envision the decay $B\rightarrow
\Sigma^+\Sigma^-$ or $\Lambda\Lambda$ .  However, a simple estimate
of the rate gives branching ratios (assuming the B-violating
couplings are unity and the scalar quark masses are near the W
mass) of
$O(10^{-8})$.   The smallness of the rate is due in large part
to two small CKM elements in the amplitude. Thus, such processes
will not provide interesting bounds unless
$10^{10}$ B-decays are studied.

It is generally assumed that the presence of both lepton
nonconserving terms and baryon nonconserving terms leads to
unacceptably rapid proton decay.  However, if enough third
generation fields are involved, proton decay can be sufficiently
suppressed as to make some of the bounds very weak (or, in a few
cases, nonexistent).  To see this, suppose both $\lm$ and $\lm''$
terms exist.  Consider the bound on the product
$|\lm_{\mu\tau\tau}\lusd|$.    Although there is a suppression due to
mixing angles and heavy squark propagators, the proton lifetime
bound gives a strong bound of $|\lm_{\mu\tau\tau}\lusd| < 10^{-14}$.
This bound is independent of the final state leptons, and thus
applies to all 9 of the $\lm$ couplings.  Similar bounds can be
obtained for all $\lm''$ couplings with at most one heavy field
(which is then the scalar quark); we obtain $|\lm_{ijk}\lubd| <
10^{-13}$,
$|\lm_{ijk}\lubs| < 10^{-12}$, $|\lm_{ijk}\lcsd| < 10^{-13}$ and
$|\lm_{ijk}\ltsd| < 10^{-12}$.  However, if the $\lm''$ coupling has
two heavy fields, a loop is necessary.  This
gives much weaker bounds; we obtain $|\lm_{ijk}\ltbs| < 10^{-2}$,
$|\lm_{ijk}\ltbd| < 10^{-3}$,
$|\lm_{ijk}\lcbs| < 10^{-3}$ and $|\lm_{ijk}\lcbd| < 10^{-2}$.  We
thus see that the lack of obvservation of proton decay does NOT
always give very strong bounds on the product of the lepton number
violating and baryon number violating couplings.

Finally, we consider the product of $\lm'$ and $\lm''$ couplings.
Here, there are $27\times 9$ possible products, of the form
$|\lm'_{ijk}\lm''_{abc}|$.   We have examined all posssible
products of couplings and found that the vast majority are tightly
bounded (product is less than
$10^{-6}$), but some are not.  Rather than list the bounds for all
243 combinations, only the bounds which are greater than $10^{-6}$
(for the product of the $\lm'$ and $\lm''$ couplings) will be
given explicitly.  It is found that all products with a $\lusd$,
$\lubd$ and $\lubs$ are smaller than $10^{-9}$.  The same is true
for $\lcsd$, $\lcbd$ and $\lcbs$, {\it except} for
$|\lm'_{tbl}\lcsd|$ which is $ < 10^{-1}$, $|\lm'_{usl}\lcbd|$
which is  $ < 10^{-2}$ and
$|\lm'_{udl}\lcbs|$, for which no bound better than the unitarity
bound could be found.  Here ${\it l}$ is any lepton.  For $\ltsd$,
the only bound which is not very small is the combination
$|\lm'_{cbl}\ltsd|$, which is bounded only by unitarity.  For
$\ltbd$ and $\ltbs$, we find $|\lm'_{ud(e,\mu)}\ltbd| < 10^{-2}$,
$|\lm'_{us(e,\mu)}\ltbs| < 10^{-2}$,
$|\lm'_{c(d,s)(e,\mu)}\ltbd| < 10^{-3}$,
$|\lm'_{c(d,s)(e,\mu)}\ltbs| < 10^{-3}$, $|\lm'_{cd\tau}\ltbd| <
10^{-5}$, $|\lm'_{cs\tau}\ltbs| < 10^{-5}$, whereas
$|\lm'_{(u,c)d\tau}\ltbs|$ and
$|\lm'_{(u,c)s\tau}\ltbd|$ are bounded only by unitarity.  All other
bounds are quite tiny.  Thus, of the 243 combinations of couplings,
thirty have bounds greater than $10^{-6}$, and
eight are completely unconstrained by the lack of observation so
far of proton decay.

It is interesting that the standard supersymmetric model can
contain some baryon number and lepton number violating coupling
constants which are of order unity, and which do not lead to
excessively fast proton decay.  Such couplings could be
measured when squarks and sleptons are discovered, since they will
lead to baryon and lepton number violating squark and slepton
decays.  These have been extensively discussed by Dreiner and
Ross and by Dimopoulos, et al.~\cite{herbi}, who analyze the impact
of such decays on phenomenology.

 \def\prd#1#2#3{{\rm Phys. ~Rev. ~}{\bf D#1} (19#2) #3}
\def\plb#1#2#3{{\rm Phys. ~Lett. ~}{\bf B#1} (19#2) #3}
\def\npb#1#2#3{{\rm Nucl. ~Phys. ~}{\bf B#1} (19#2) #3}
\def\prl#1#2#3{{\rm Phys. ~Rev. ~Lett. ~}{\bf #1} (19#2) #3}

\bibliographystyle{unsrt}

\end{document}